
\input harvmac

\def \s {\sigma}

\def \G {\Gamma}

\def \p {\phi}
\def \ha {\half}
\def \ov {\over}

\def \four {\textstyle {1\ov 4}}
\def \a {\alpha}
\def \lr { \lref}
\def\ep{\epsilon}

\def\vp {\varphi}
\def \bd {\bar \del}
\def \r {\rho}
\def\bd {\bar \del} \def\m{\mu}\def\n {\nu}\def\l
{\lambda}\def \G {\Gamma}

\def \four { {\textstyle{1 \ov 4}}}

\def \y { \tilde \y}

\def \g {\gamma}

\def\y {{ \tilde y}}

\def   \td {\tilde }

\def \lr { \lref}

\gdef \jnl#1, #2, #3, 1#4#5#6{ { #1~}{ #2} (1#4#5#6) #3}

\def\np {  Nucl. Phys. }
\def \pl { Phys. Lett. }
\def \mpl { Mod. Phys. Lett. }
\def \prl { Phys. Rev. Lett. }
\def \pr  { Phys. Rev. }

\def \cmp  { Commun. Math. Phys. }

\def \ijmp {Int. J. Mod. Phys. }

\def \eg  {{\cal G }}
\def \eh  {{\cal H }}
\def \el  {{\cal L }}
\baselineskip8pt
\Title{\vbox
{\baselineskip 6pt{\hbox{ }}{\hbox
{Imperial/TP/94-95/34 }}{\hbox{hep-th/9505129}} {\hbox{  }}} }
{\vbox{\centerline { On gauge theories for non-semisimple groups}}}

\vskip -3 true pt

\centerline{   A.A. Tseytlin\footnote{$^{\star}$}{\baselineskip8pt
e-mail address: tseytlin@ic.ac.uk}\footnote{$^{\dagger}$}{\baselineskip8pt
On leave  from Lebedev  Physics
Institute, Moscow, Russia.} }

\smallskip\smallskip
\centerline {\it  Theoretical Physics Group, Blackett Laboratory}
\smallskip

\centerline {\it  Imperial College,  London SW7 2BZ, U.K. }
\bigskip\medskip
\centerline {\bf Abstract}
\medskip
\baselineskip8pt
\noindent
We consider analogs of Yang-Mills theories for non-semisimple
real Lie algebras which admit invariant non-degenerate metrics.
These 4-dimensional theories have  many similarities with
corresponding  WZW  models in 2 dimensions and Chern-Simons
theories in 3 dimensions. In particular, the  quantum effective
action contains only 1-loop term with the divergent part that
can be eliminated by a field redefinition. The on-shell scattering
amplitudes are thus finite (scale invariant). This is a consequence
of the presence of a null direction in the field space metric:
one of the field components is a Lagrange multiplier
which `freezes out' quantum  fluctuations  of  the `conjugate' field.
The non-positivity of the metric implies  that these theories
are apparently  non-unitary. However, the  special structure of
interaction terms (degenerate compared to non-compact YM theories)
suggests that there  may exist a  unitary `truncation'.
We discuss in detail the simplest theory based on  4-dimensional
algebra $E^c_2$. The quantum part of its effective action is
expressed in terms of 1-loop effective action of $SU(2)$ gauge theory.
The $E^c_2$ model can be  also described  as a special limit
of $SU(2) \times  U(1)$ YM theory with decoupled ghost-like $U(1)$ field.

\Date {May  1995}

\noblackbox
\baselineskip 14pt plus 2pt minus 2pt

%


\lr \lee {T.D. Lee and G.C. Wick, \np B9 (1969) 209; \pr D3 (1970) 1046;
D.G. Boulware and D.J. Gross, \np B233 (1984) 1.}

\lr \mast {  M.B. Halpern and E.B. Kiritsis, \mpl A4 (1989) 1373; 1797
(E);
A.Yu. Morozov, A.M. Perelomov, A.A. Rosly, M.A. Shifman and A.V.
Turbiner, \ijmp
A5 (1990) 803;
 M.B. Halpern, E. Kiritsis, N.A. Obers and K. Clubok, ``Irrational conformal
field theory", hep-th/9501144.}

\lr \wein {S.  Weinberg,  in: General Relativity, eds. S.Hawking and W. Israel
(Cambridge Univ. Press, 1979).}

\lref \napwi { C. Nappi and E. Witten, \jnl \pl,  B293, 309, 1992.}

\lr \oliv{ D.  Olive,
 E. Rabinovici and A. Schwimmer,  \jnl \pl, B321, 361, 1994.}
\lr\sfetso{ K. Sfetsos,  \jnl \pl, B324, 335, 1994;
\ijmp A9 (1994) 4759.}
 \lr \sf{ K. Sfetsos, \jnl \pr,  D50, 2784,  1994.}

\lr \hrts { G. Horowitz and A.A. Tseytlin, \jnl \pr, D51, 2896, 1995.}

\lr \moh { N. Mohammedi, \jnl \pl,  B325, 379,  1994. }

\lr\figu{
 J.M. Figueroa-O'Farrill and S. Stanciu, \jnl \pl, B327, 40,  1994. }

\lr \keha{
A.  Kehagias and P.  Meessen, \jnl  \pl,  B331, 77, 1984;
A.  Kehagias, ``All WZW models in $D\leq 5$", hep-th/9406136.}

\lr \kz {V.G. Knizhnik and A.B. Zamolodchikov, \np B247 (1984) 83. }

\lr \sfeets { K. Sfetsos and A.A. Tseytlin, \jnl \np, B427, 245, 1994. }

\lr \kk {E. Kiritsis and C. Kounnas, \jnl \pl, B320, 264, 1994;
E. Kiritsis, C. Kounnas and  D. L\"ust, \jnl \pl,  B331, 321,  1994.}
\lr \bars {I. Bars, ``Ghost-free spectrum of a quantum string in $SL(2,R)$
curved spacetime", hep-th/9503205.}
\lr \jac{D. Cangemi and R. Jackiw, Phys. Rev. Lett. { 69} (1992) 233;  Ann.
Phys. (NY) { 225} (1993) 229.}
\def \th {\theta}

\def \O {\Omega}
\lr \dewit{B.S. DeWitt, Dynamical Theory of Groups and Fields (Gordon and
Breach, 1965). }
\lr \hooft{G. 't Hooft, \np B62 (1973) 444. }
\lr \palla{ }

\def \F {{\tilde F}}
\def \A {{\tilde A}}

\def \g {\gamma}

\lr \sib { G. Moore and N. Seiberg, \pl B320 (1989) 422. }

\lr \leut {H. Leutwyler and M. Shifman, \ijmp A7 (1992) 795.}

\lr \ichi {L.F. Abbott, \np B185 (1981) 189;
S. Ichinose and M. Omote, \np B203 (1982) 221.}
\lr \gro{H.D. Politzer, \prl 30 (1973) 1346;
D.J. Gross and F. Wilczek, \prl 30 (1973) 1343.}

\lr \wiit{E. Witten, \cmp 121 (1989) 351.}
\lr \wit{E. Witten, \cmp 92 (1984) 455.}
\lr\jad{S. Deser, R. Jackiw and S. Templeton, \prl 48 (1982) 975;
 Ann. Phys. 140 (1982) 372. }
\lr \palla { P. Forg\' acs, P.A. Horv\' athy, Z. Horv\' ath and L. Palla,
``The Nappi-Witten string in the light-cone gauge", hep-th/9503222.}

\lr \schw {A.S. Schwarz, Lett. Math. Phys. 2 (1978) 2247.}
\lr \shif{M.A. Shifman, \np B352 (1991) 87.}

\lr \ferm {A.N. Redlich, \prl 52 (1984) 16;
A.J. Niemi and G.W. Semenoff, \prl 51 (1983) 2077;
R. Pisarski  and S. Rao, \pr D32 (1985) 2081.}
\lr \others{
L. Alvarez-Gaum\' e, J.M.F. Labastida and A.V. Ramallo, \np B334 (1990) 103;
I.I. Kogan, preprint ITEP - 89 -163 (1989); Comm. Nucl. Part. Phys.  19 (1990)
305; I.I. Kogan and V. Fock,  Mod. Phys. Lett. A5 (1990) 1365;
G. Giavarini, C.P. Martin and F. Ruiz Ruiz, \np B381 (1992) 222.
}

\lr \tse {A.A. Tseytlin, \np  B399 (1993) 601; B411 (1994) 509.}
\lr \gri { B. de Wit, M.T. Grisaru and P. van Nieuwenhuizen,
\np B408 (1993) 299.}

\lr \casw{D.R.T. Jones, \np B75 (1974) 531; W.E. Caswell, \prl 33 (1974) 244.}

\lr \rts{ J.G.  Russo and A.A. Tseytlin, ``Constant magnetic field in closed
string theory: an exactly solvable model", CERN-TH.7494/94,
Imperial/TP/94-95/3, hep-th/9411099. }

\def \b {\beta}
\def \g {\gamma}
\def \d {\delta}

\lr\kli{ C. Klim\v c\'\i k and A.A. Tseytlin, \jnl \pl, B323, 305, 1994.}

\lr\anto{I. Antoniadis and N. Obers, \jnl \np, B423, 639, 1994.}

\newsec{Introduction}
In contrast to 2 dimensions
the number of
solvable or conformal  models in 4 dimensions  seems to be  very limited.
The only known examples are special  Yang-Mills  theories
with (extended) supersymmetry.
It may be of interest to  study  bosonic  Yang-Mills -type  $D=4$ models which
are  analogs of   certain  2-dimensional models and have  simpler quantum
properties than standard non-abelian
gauge theories.

 Many conformal field
theories in two dimensions are described by (gauged) WZW models.
It was realized in  \napwi\  that  one can also construct WZW models for some
non-semisimple groups (which admit a nondegenerate invariant bilinear form)
and such models   \refs{\oliv, \kk,\sfetso,\moh,\figu,\sf,\anto,\sfeets,\keha}
  have much simpler structure  than  the standard semisimple models.  One
possible explanation
 for that is that they can be interpreted as formal
limits of semisimple coset or WZW models where one makes simultaneous  boost of
 coordinates and  infinite rescaling of the  level
\refs{\oliv,\sfetso,\sfeets}.

There is a  certain analogy between semisimple WZW  theory   in 2 dimensions
and    Yang-Mills  theory in 4 dimensions.
Both  are uniquely   defined by specifying a  Lie group and
 are  classically conformally invariant.
However, in contrast to WZW model,
 Yang-Mills  theory  is no longer  conformal  at the quantum level
and has complicated  dynamics.
Below we shall  construct   Yang-Mills  theories  based on some non-semisimple
groups  and  demonstrate that like their WZW counterparts
they turn out to be very simple  (though not exactly conformal)  also at the
quantum level.

The main difference compared to 2-dimensional WZW case is an apparent lack of
unitarity  of
these   non-semisimple 4-dimensional  models.
While
in  two  dimensions
the presence of one  negative norm direction in the non-degenerate  invariant
bilinear  form which replaces the  Killing form  may  not be, in principle, a
problem since  there is  an
infinite dimensional conformal symmetry and  hence   (as in flat space case)
we may be able to   eliminate  the `time-like' field component  by choosing
the light-cone  gauge (which indeed exists  in these  WZW models),\foot{The
issue of unitarity of WZW models based on non-compact semisimple groups where
one is unable to  choose  a light-cone
 is  nontrivial  and   remains under discussion (see \bars\ a recent
suggestion of a unitary formulation of the model and  also refs. there).}
  this is no longer possible
in four dimensions.
However, the non-unitarity of  the non-semisimple Yang-Mills  models  is much
`milder' than that of  Yang-Mills  theories  based on non-compact simple
groups.
For example,
the special `null' structure of the interaction terms implies that
one linear combination of the positive and negative norm fields is
  decoupled.

We shall start in Section 2 with a review of
the structure on non-semisimple real Lie algebras with invariant non-degenerate
bilinear forms
following \figu.
All explicitly discussed examples of such algebras
\refs{\jac,\napwi,\oliv,\sfetso,\sf}
are so called  indecomposable `depth 1' algebras  \figu\ which are `double
extensions'  of
an abelian algebra by a simple or 1-dimensional one.

Given a Lie algebra with an invariant  metric $\O_{ab}$
it is straightforward to define the non-semisimple
generalizations of the WZW \napwi, Chern-Simons and Yang-Mills  actions
(Section 3). If  $\O_{ab}$ is put  into a diagonal form
the actions have the same structure as in the case of a simple noncompact
algebra but with  `degenerate' structure constants.
Since it is  the structure constants that determine the interactions,
the quantum correction  to  $\O_{ab}$  in WZW and CS  theories  (represented
exactly by the 1-loop term) is proportional  to
the degenerate Killing metric $g_{ab}$.

Similar  renormalization of $\O_{ab}$  is found in  non-semisimple Yang-Mills
case (Section 4):
$\O_{ab} \to \O_{ab} + b_1 g_{ab}$, where now  $b_1$ is not finite but
logarithmically divergent. The analogy with WZW and CS models goes beyond 1
loop:
in contrast to semisimple Yang-Mills  theories,  in the non-semisimple case
there is no two and higher loop  renormalization.
 We demonstrate this for
 `depth 1'  algebras  but this is   likely  to be true for  generic
 indecomposable non-semisimple algebras with invariant metrics.

Moreover,
as in the non-semisimple WZW and CS models,
 the full quantum effective action is  given just by the 1-loop term.
The divergent part of the effective action can actually   be eliminated by a
field redefinition (i.e. the  infinite renormalization  mentioned above
is an off-shell artifact).
  This implies
that the on-shell  S-matrix is UV finite.\foot{Note that in contrast to, e.g.,
higher derivative or Pauli-Villars regularized theories which are also finite
at the expense of being  superficially non-unitary \lee,  here  one  has
no extra scale-dependent parameters.}
The theory is  very similar to  that of a  collection  of
 quantized  (abelian) positive norm vector fields
interacting  only  with a   background  gauge field
subject to the classical Yang-Mills  equation.
In particular, the conformal anomaly depends only on the background gauge
field.
This suggests a possibility of a (unitary, conformal)  `free-field'
interpretation of this model.

The general discussion  will be illustrated in  Section 5
on the simplest nontrivial example of the  4-dimensional algebra
$E^c_2$.   We shall first review the  structure of the
$E^c_2$ WZW  theory
and then consider the $E^c_2$ Yang-Mills  theory,
making  the analogies with the  $E^c_2$ WZW and  CS models explicit.
We shall  also  describe   the
 relation  between the  $E^c_2$ theory  and  the $SU(2) \times U(1)_-$
Yang-Mills   theory (with decoupled $U(1)$ ghost field).

Our conclusions will be summarized in Section 6.

\newsec{Non-semisimple Lie algebras with invariant metrics}
For  a real
 Lie algebra with generators $e_a$  ($a=1,...,N$) and structure constants
$f^a_{\ bc } $  there   may exist  a symmetric, invariant and non-degenerate
bilinear form $\O_{ab}$  which generalizes the Killing form $g_{ab}$
\eqn\yyme{  [e_a, e_b] = f^c_{\  ab} e_c\ ,  \ \ \  \ \ g_{ab} \equiv
 - f^c_{\  ad}f^d_{\  bc} \ , \ \ \ \ g_{c(d}f^c_{\  a)b}=0  \ ,  }
\eqn\yym{ \Omega_{ab}=\Omega_{ba}\ , \ \ \ \
\Omega_{c(d}f^c_{\  a)b}=0  \ , \ \ \ \   \det \Omega_{ab} \not=0 \ , \ \
\O^{ab}\equiv  (\O\inv)^{ab}\ .   }
In the semisimple case $g_{ab} = c_2 \eta_{ab} , \
 \ \O_{ab} =  \g g_{ab} = k  \eta_{ab}  $, where $\g$ is a constant, $k= c_2
\g$  and $\eta_{ab}$ is diagonal matrix with $\pm 1$ entries.
In  general,  we may  always represent $\O_{ab}$
in the form
\eqn\defqw{  \O_{ab}=\g g_{ab} + \O^{(0)}_{ab}   \ , }
where $\O^{(0)}_{ab}$
  also satisfies
the invariance condition  in \yym\  and can be considered
a `complement' of  $g_{ab}$  which makes
$ \O_{ab}$ non-degenerate.
Thus  generic $\O_{ab}$ contains at least
one free parameter $\g$. As we shall see below, it is this parameter that is
renormalized by quantum corrections (and its renormalization is  formally the
same as in the semisimple case).

 Examples of such algebras were  considered in
\refs{\jac,\napwi,\oliv,\sfetso,\sf,\keha}.\foot{The first nontrivial
non-semisimple examples are $N=4$
ones: $E^c_2$ and Heisenberg algebra $H_4$
 (which have the same complex extension).}
The general construction was  presented in \figu\  and will be reviewed below.
According to a theorem discussed in \figu\
the class of real Lie algebras with an invariant metric is the product of the
class of semisimple algebras  with the class
obtained from the 1-dimensional algebra under the operations of taking direct
sum and double extension by  simple (or 1-dimensional) algebras.
Indecomposable (i.e. not equal to a direct product)
non-semisimple  dimension $>1$ Lie algebra  $\el$  with  an invariant metric
is always  a double extension  of another Lie algebra $\eg$ with an invariant
metric $\O^{(0)}_{ij}$  by a Lie algebra  $\eh$ which is either simple or
1-dimensional.
Equivalently, $\el$ is  a semidirect
sum of $\cal H $ with  ${\cal G} \oplus_c {\cal H}^*$, i.e.   whith  a central
extension of $\eg$ by the abelian algebra $\eh^*$
($\eh^*$ is dual of $\eh$ and is  an abelian ideal of  $\el$).  What that means
is that
 $\el= \eg \oplus \eh \oplus \eh^*$ as a vector space,
and the commutation relations in
 the    basis $e_a =\{e_i, e_r, e^s\}$  ($i,j,k,l=1,..., \dim \eg, \ \
r,s,t= 1,...,\dim \eh=\dim \eh^*$) are
 \eqn\me{  [e_i, e_j] = f^k_{\  ij} e_k  +  f_{i jr} e^r \ , \ \ \
[e_r, e_s] = f^t_{\  rs} e_t \ , }
$$ [e_r, e_i] = f^j_{\  ri} e_j  \ , \ \ \
[e_r, e^s] = f^s_{\ tr  } e^t \ , \ \ \ [e^r, e^s] = 0 \ ,
\ \  [e^r, e_i] = 0  \ , $$
$$ f_{i jr}\equiv \O^{(0)}_{ik} f^k_{\ jr} \ , \ \ \
  f_{(k j)r}=0 \ , \ \  \  \
f^k_{\ [ij } f^l_{ \  r]k} =0 \ . $$
The corresponding Killing form is degenerate
\eqn\kill{
g_{ab}=  f^c_{\  ad}f^d_{\  cb} = \pmatrix{g_{ij} & g_{is} & 0 \cr
g_{rj} & g'_{rs} & 0 \cr
0 & 0 & 0\cr}\
 \ ,  }
$$  \ \   g_{ij}=  f^k_{\  il}f^l_{\  kj} \ ,
\ \
 g_{ir}= g_{ri}= f^k_{\  il}f^l_{\  kr} \ , \ \  \
g'_{rs} =  f^i_{\  rj}f^j_{\  is} + 2 g_{rs}\ , \ \ \    g_{rs}=  f^t_{\  rp}
f^p_{\  t s} \ ,  $$
but there exists  a non-degenerate invariant form
\eqn\invm{
\O^{(0)}_{ab}=\pmatrix{\O^{(0)}_{ij} & 0 & 0 \cr
0 & \O^{(0)}_{rs} & \delta^s_r \cr
0 & \delta^r_s  & 0\cr}\
 \ ,  }
where $\O^{(0)}_{rs}$ is an arbitrary
 (possibly degenerate) invariant bilinear form on $\eh$.
For a given $\el$ \me\ the free parameters in
$\O^{(0)}_{ab}$   are at least the $\O^{(0)}_{rs}$ and the   scale of
$\O^{(0)}_{ij}$.
Note that the signature of $\O^{(0)}_{ab}$ is always indefinite.
More general $\O_{ab}$ in \defqw \ is then given by
\eqn\defqe{  \O_{ab}=\g g_{ab} + \O^{(0)}_{ab}
=\pmatrix{\O_{ij} & g_{is} & 0\cr
g_{rj} & \O_{rs} & \delta^s_r  \cr
0 &  \delta^r_s  & 0\cr}\ ,  } $$ \ \O_{ij}= \g g_{ij} + \O^{(0)}_{ij} \ ,
\ \ \ \ \  \O_{rs}= \g g'_{rs}  + \O^{(0)}_{rs} \ .
$$
According to \figu\
$\cal G$ itself must  be   a double extension
of   some other algebra ${\cal G}_1$ with an invariant metric $\O^{(0)}_1$
by some simple or 1-dimensional algebra ${\cal H}_1$,
etc.  After a number of double extension steps
the final algebra  ${\cal G}_n$ must be abelian.
All  previously  explicitly discussed examples
\refs{\napwi,\oliv,\sfetso,\sf,\keha}  correspond to a {\it single}
 double extension
 of some {\it abelian}  Lie algebra $\eg$, i.e. can be called
 `depth 1' algebras \figu.
 In  the general case
 of  a `depth 1' indecomposable
non-semisimple $\cal L$  with   simple or 1-dimensional $\cal H$  and
abelian  $\eg$ ($f^i_{\ jk}=0$)
the Killing metric \kill\  has  $g'_{rs}$ as the only non-vanishing entry
($g_{ij}=0, \ g_{ir}=0$)
and  $ \O_{ab}$ \defqe\  takes  the same form as $\O^{(0)}_{ab}$ \invm,
\eqn\defe{  \O_{ab}
=\pmatrix{\O^{(0)}_{ij} & 0 & 0\cr
0 & \O_{rs} & \delta^s_r  \cr
0 &  \delta^r_s  & 0\cr}\ ,  \
\ \ \   g_{ab}=\pmatrix{0 & 0 & 0 \cr
0 & g'_{rs} & 0 \cr
0 & 0 & 0\cr}\
 \  , }
\eqn\defr{   \O^{ab}=\pmatrix{\O^{(0)ij} & 0 & 0 \cr
0 & 0 & \delta^r_s \cr
0 & \delta^s_r  & - \O_{rs} \cr}\  ,   \ \ \ \    \O_{rs} = \g g'_{rs}  +
\O^{(0)}_{rs} \ . }
Note that since each double extension step  adds an equal number
 ($\dim {\cal H}$)  of pluses
and  minuses  to the signature of $\O_{ab},$
to get the  signature with just one minus
(to `minimize' the non-unitarity of  a   field theory
constructed using $\O_{ab}$)
one should consider only `depth 1' algebras
which  have 1-dimensional $\cal H$.\foot{Double extensions of
semisimple $\cal G$ by  1-dimensional $\cal H$
is decomposable with simple part of  $\cal G$ as a factor \figu.}
The only such algebras
are  the  trivial  $N$-dimensional generalization of $E^c_2$ \sf\
  or the Heisenberg algebra  $H_N$ \keha\ (see Section 5).

\newsec{Wess-Zumino-Witten, Chern-Simons and Yang-Mills theories
for non-semisimple  groups}
\subsec{Classical actions}

Given a Lie algebra  $\cal L$  with an invariant metric one may  define the WZW
model \wit\  by the following action \napwi\
\eqn\wzz{ I = { 1\ov 16\pi}\O_{ab}\big( \int_{\del M} d^2 \s \  A^a_\m A^b_\m
+ {i\ov  3  } \int_{M}  d^3 x\   \ep^{\m\n\l} f^a_{\  cd }  A^c_\m A^d_\n
A^b_\l \big)  \ , }
where  2-metric is flat and $A^a_\m$ is defined by
\eqn\deff{ A^a_\m e_a = g\inv \del_\m g \ , \ \ \  \  \ g=\exp (x^a e_a) \ . }
Similarly, we can  construct  the
`non-semisimple' generalizations
of the $D=3$ Chern-Simons  action \schw
\eqn\cher{ S_{CS}= - { i\ov 8\pi}\ \O_{ab}  \int  d^3 x \   \ep^{\m\n\l}
\big( A^a_\m \del_\n  A^b_\l + {1\ov 3}  f^a_{\  cd }  A^c_\m A^d_\n
A^b_\l\big) \ , }
and of the $D=4$ Yang-Mills action
\eqn\yyym { S_{YM} = {1\ov 4}\Omega_{ab}  \int d^4x \   F^a_{\m\n}  F^b_{\m\n}
\ , }
\eqn\fef{\ F^a_{\m\n} = 2\del_{[\m} A^a_{\n]}  + f^a_{\  bc} A^b_\m A^c_\n \  ,
}
which are invariant under
\eqn\gau{\delta A^a_\m = \del_\m \eta^a + f^a_{\ bc} A^b_\m \eta^c\  ,  \ \
\ \ \ \delta F^a_{\m\n} = f^a_{\ bc} F^b_{\m\n} \ep^c   \ . }
On can also consider other  actions, e.g., a combination
of \cher\ with $D=3$ analog of \yyym\
(i.e. the topologically massive $D=3$ gauge theory \jad)
or the straightforward generalization of \yyym\
including the topological term $\theta \O_{ab} \int F^a_{\m\n} \td F^b_{\m\n}$.
The theories \wzz\  and \cher\ are  closely related as in the semisimple case
\refs{\wiit,\sib} (the classical equation that follows from \cher\ is still
$F^a_{\m\n}=0$).
In  \wzz,\cher,\yyym\
 one may  separate the overall loop counting  parameter (`Planck constant') by
setting
$\O_{ab} = k \bar \O_{ab} , \  k = 1/g^2. $

Since $\O_{ab}$ is non-degenerate,
it can be  represented  in the `diagonal' form
\eqn\dii{ \O_{ab} = E^\a_a E^\b_b \eta_{\a\b}\ , \ \ \  \
\eta_{\a\b} = {\rm diag} (+1,...,+1, - 1, ..., -1)\ , \ \ \ \det E^\a_a \not=0
\ . }
One can  then     define
the `rotated' structure constants  with   `vierbein' indices ($\a,\b,\g$) which
will  encode the information about  $\O_{ab}$
\eqn\feff{ f^\a_{\ \b\g} \equiv  E^\a_a E^b_\b E^c_\g  f^a_{\ bc} \ ,
\ \ \ \  \ f_{\a\b\g} \equiv \eta_{\a\d} f^\d_{\ \b\g} \ ,
} $$     f_{(\a \b)\g} =0  \ , \ \ \
\ \ \ g_{\a\b}
=  - f^\g_{\ \a\d}f^\d_{\  \b\g}= E^a_\a E^b_\b g_{a b}  \ .  $$
Introducing  the `rotated' gauge field $A^\a_\m \equiv E^\a_a A^a_\m$
one can  put the actions \cher\ and \yyym\ in the equivalent forms
\eqn\cheri{ S_{CS}= - { i\ov 8\pi}\ \eta_{\a\b}  \int  d^3 x \   \ep^{\m\n\l}
\big( A^\a_\m \del_\n  A^\b_\l + {1\ov 3}  f^\a_{\  \g \d }  A^\g_\m A^\d_\n
A^\b_\l\big) \ , }
\eqn\yyymi { S_{YM} = {1\ov 4}\eta_{\a\b}  \int d^4x \   F^\a_{\m\n}
F^\b_{\m\n} \ ,  \ \ \  F^\a_{\m\n} = 2\del_{[\m} A^\a_{\n]}  + f^\a_{\  \b\g}
A^\b_\m A^\g_\n \  . }
These  actions look the same as in the case of a semisimple non-compact group
but  with the structure constants corresponding to the non-semisimple
algebra \yyme\ in the `rotated' basis $e_\a = E^a_\a e_a$.
Since  $\eta_{\a\b}$ is not positive definite an apparently obvious conclusion
is that
 the  YM theory \yyymi\ is non-unitary. Indeed, the Hamiltonian or Euclidean
action are not positive.
However, this non-unitarity may be less serious  than in the case of a
non-compact simple group  due to   the  `degenerate' form of the structure
constants which determine the interactions (for example,  ghost-like gauge
vectors  may effectively  decouple and one may be able to separate them from a
unitary sector of the theory).

\subsec{Quantum renormalization of couplings: WZW and CS theories}
Let us now discuss the corresponding quantum theories.
It is easy to show  that   WZW model \wzz\  remains conformal for {\it any}
 $\O_{ab}$
in \yym. There is a finite quantum shift of $\O_{ab}$
which can be determined, e.g.,  by generalizing the standard
semisimple current algebra approach \kz, or, equivalently,
by solving the master equation \mast.\foot{Our
 discussion (which is a generalization  of that of \napwi)
is slightly different  from  that of \refs{\moh,\figu}
in that we start directly with the action \wzz\ which fixes
the form of the current algebra.}
Starting with the action \wzz, defining the currents
$J_a= \O_{ab}A^b_z$ and imposing the canonical commutation relations  one finds
 that $J_a$  generate
the  affine algebra with structure constants $f^a_{\ bc}$ and
the central term proportional to $\O_{ab}$.
 The  quantum analog of the classical stress
tensor  $T_{zz} =  \O_{ab} A^a_z A^b_z= \O^{ab}J_a J_b$
is  $\hat T_{zz} =  {\hat \O}^{ab}:J_a J_b:$,  \ where
${\hat \O}^{ab}$  should satisfy \mast\foot{The relation to the notation in
\mast\ is
${\hat \O}^{ab}=L^{ab}, \ \O_{ab} = 2G_{ab}\to  2k_I\eta_{ab}, \  k=2k_I. $ }
\eqn\mases{
 {{\hat \O}}^{ab} =   {  {\hat \O}}^{ac} \O_{cd} { {\hat \O}}^{db}
  - { {\hat \O}}^{cd}{{\hat \O}}^{ef}
  f^a_{\ ce} f^{b}_{\ df}
 - 2{ {\hat \O}}^{cd}f^{f}_{\ ce} f^{(a}_{\ \ df} { {\hat \O}}^{b)e}
 \  . }
Assuming that ${\hat \O}^{ab}$ is non-degenerate
and multiplying \mases\ by ${\hat \O}_{aa'}{\hat \O}_{bb'}, \ {\hat
\O}_{ab}\equiv ({\hat \O}\inv)_{ab}$
one finds (using \yym)  the following solution
\eqn\mass { {\hat \O}_{ab} =  \O_{ab} +  g_{ab} \ . }
The corresponding central charge of the Virasoro algebra
is
\eqn\cen{ c = \O_{ab} {\hat \O}^{ab} = \Tr\big({\O\ov \O +  g}\big)  \ . }
As can be shown using \me--\invm\  $c$   is always integer  for
indecomposable   non-semisimple $\cal L$ \figu.
These relations are the generalizations of the standard expressions in the
semisimple
 case ${\hat \O}_{ab} = (k +  c_2)\eta_{ab}, \ \  c = Nk/(k + c_2)$.
The same results can be reproduced in the Lagrangian field theory approach,
e.g., by
generalizing the discussion in
\refs{\leut,\tse,\gri}.
  In particular, the  local part of the quantum effective action   will be
given by \wzz\ with $\O_{ab}$ replaced by $\hat \O_{ab}$.

The  important
observation  that will be true also in CS and YM theories is that
while the  classical action depends on $\O_{ab}$,  the   quantum  correction
contains  only the  contraction of the two  structure constants, i.e. is
proportional to  the  (degenerate) Killing metric.
Thus the  part of $\O_{ab}$  which  is not proportional to $g_{ab}$
is {\it not renormalized}, i.e.
the only effect of renormalization is  to shift the constant $\g$ in \defqw,
\eqn\defq{ \hat \O_{ab}= \hat \g g_{ab}  + \O_{ab}^{(0)}\ , \ \ \  \ \
\ \hat \g = \g + 1 \  .  }
The   reason underlying  this conclusion  is that the  relevant quantum
correction is exactly given by the
1-loop diagram with two propagators (containing $\O\inv$)
and two vertices (proportional to $\O_{ab}f^b_{\ cd}$)
so that the  $\O$-factors
cancel out because of the invariance property \yym.

Similar quantum shift of $k$  is known to occur in the semisimple  Chern-Simons
theory
\wiit\ (see also, e.g.,  \refs{\others,\shif}). It is straightforward to
generalize, e.g.,  the perturbative background field method  derivation of the
induced
P-odd  (CS) term  in the effective action
 \shif\  to the case of the action \cher.\foot{The  off-shell
effective action in general contains other P-even terms but the
shift of $k$ that comes from the 1-loop determinant
is the only genuine `on-shell' effect \refs{\wiit,\shif}.
This  shift, of course, depends on the definition of the determinant,
i.e. on a  scheme choice.}
Starting with \cheri\  and splitting the  gauge field  into the   background
$A^a_\m$  and quantum $ B^a_\m $ parts one can represent the term in the action
which is bilinear in $B$ and  auxiliary gauge fixing  field $\p$ in the
form\foot{The corresponding ghost term can be ignored since it does not
contribute to the P-odd part of the  effective action.}
\eqn\cherr{ S^{(2)} = - { i\ov 8\pi} \eta _{\a\b}  \int  d^3 x  \big(
\ep^{\m\n\l}
B^\a_\m D_\n   B^\b_\l  + 2\p^\a D_\m  B^\b_\m \big) \ , }$$   \ \ \ (D_\m
)^\a_\g \equiv \delta^\a_\g \del_\m + f^\a_{\ b\g } A^b_\m  \  ,  $$
which can be then be represented as  a  fermionic action,
\eqn\chee{- { i\ov 8\pi}\ \eta_{\a\b}  \int   d^3 x \  \psi^\a \Sigma^\m D_\m
 \psi^\b \ , }
where $\psi = (B_\m, \p)$ and $\Sigma^\m$ are 4$\times$4 matrices with
properties similar to those of  Pauli matrices \shif.
The resulting quantum contribution
 looks  the same as in the semisimple case  \refs{\ferm,\wiit,\others,\shif},
i.e.
is given by  the classical action \cher\  with $\O_{ab}$ replaced by the
contraction of the two structure constants $ f^\a_{\ a \b} f^\b_{\ \a b } , $
i.e. by  the Killing metric $g_{ab}$.
As a consequence, the quantum
effective action  contains  the same shifted $\O_{ab}$  as in
\mass,\defq.

\newsec{Non-semisimple Yang-Mills theory at the quantum level}
\subsec{One-loop renormalization and absence of higher loop  divergences}
Next, let us  consider the  renormalization
of the coupling matrix $\O_{ab}$ in the  YM  theory  \yyym.
Again, one may expect that the 1-loop correction
(which  will now be infinite) will not depend on $\O_{ab}$ itself.
Using the background field method \refs{\dewit,\hooft}
one can put the relevant  `quantum'  part of the action \yyymi\
in the form (cf. \cherr)
\eqn\onee{
 S  =   \int d^4x\  \big(
 \ha \eta_{\a\b} D_{\m} B^\a_{\n } D_{\m} B^\b_{\n }
 + f_{\a\b\g}   F^\a_{\m\n} B^\b_\m B^\g_\n
 + f_{\a\b\g}   D_{\m}B^\a_\n B^\b_\m B^\g_\n  }
$$
+\  \four f_{\a \b\g} f^\a_{\ \d\s} B^\b_\m B^\g_\n B^\d_\m  B^\s_\n
+ \eta_{\a\b} D_{\m} \bar c^\a D_{\m} c^\b  + f_{\a \g\b} B^\g_\m D_{\m} c^\a
c^\b \big) \ , $$
where we have added the background gauge fixing term  $\eta_{\a\b}  D_{\m}
B^\a_{\m } D_{\n} B^\b_{\n }$ and the corresponding ghost term.  The  1-loop
correction to the effective action  $\G^{(1-loop)}$
depends on $\O_{ab}$
 through the structure constants $f^\a_{\ \b \g}$ in  \feff.
However, this dependence cancels out in the contraction of two structure
constants which appears in the divergent part of  $\G^{(1-loop)}$.
Indeed, repeating the computation in \hooft\ (see also
\ichi) one finds that the  divergent part of the 1-loop effective action
is given by formally the same expression as in the semisimple case
\refs{\hooft,\gro}
\eqn\onel{  \G^{(1-loop)}_\infty= {1 \ov  \ep }\b_1  g_{ab}  \int d^4x \
F^a_{\m\n}  F^b_{\m\n} \ ,  \ \  \ \ \b_1 = - {11\ov 6 (4 \pi)^2 } \ , \ \ \
\ep=4-D \to 0 \ . }
For a  simple  Lie algebra
 $g_{ab} = c_2 \eta_{ab}$ while  in the general case we learn that
the 1-loop effective action  contains the classical
action term  with  the tree-level coupling  matrix
$\O_{ab}$ replaced by (cf. \mass,\defq)
\eqn\reno{ \hat \O_{ab}^{(1-loop)} = \O_{ab}  +  {1 \ov  \ep }\b_1 g_{ab} \ ,
\ \ \  \ \ \hat \g^{(1-loop)} =\g +  {1 \ov  \ep }\b_1 \ . }
As in WZW and CS models   the  non-degenerate part  $\O_{ab}^{(0)}$
 of the coupling matrix  \defqw\
 which is `complementary' to the Killing form $g_{ab}$
 is not renormalized.\foot{Note that had  we   set $\g=0$ in the classical
action (which  could  seem  formally possible  since  $\O^{(0)}_{ab}$
can be chosen to be non-degenerate)
the theory \yyym\ would not be renormalizable, i.e.
 to get a renormalizable
YM theory one needs to start with the most general  $\O_{ab}$.}

 In general, the
divergent part of the  $n$-loop term in the effective action in the theory
\onee\
  should
have the form
\eqn\coun{ \G^{(n-loop)}_\infty =\sum_{m=0}^n {\b_m \ov  \ep^m } C^{(nm)}_{ab}
\int d^4x \   F^a_{\m\n}  F^b_{\m\n} \ ,   }
where the matrices $ C^{(nm)}_{ab}$ are   built out of $f^a_{\ bc} $,
 $\O_{ab}$ and  $\O^{ab}$,  or, equivalently,  out of products of $f_{\a \b\g}$
and $\eta_{\a\b}$.
 Since   $f_{\a\b\g}$ satisfies the same
properties (total antisymmetry and Jacobi identity)
as the structure constants of a  semisimple  algebra,
 higher loop contributions $C^{(nm)}_{ab}$  will also  have identically the
same form as in the
standard semisimple YM theory {\it before } one uses there
the  expressions for the products of the structure constants in terms of
$\eta_{\a\b}$ which are true only in the semisimple case,
 $ \
 f^\d_{\ \a \g} f^{\g}_{ \  \d \b} =     f_{\g  \d\a}f^{\g \d}_{\ \  \b} =  c_2
\eta_{\a\b}\ .$
Since the number of $f$'s in $C^{(nm)}_{ab}$  is always even,
and because of the properties of $f_{\a\b\g}$,  the second rank tensors built
out of them with the help of $\eta_{\a\b}$ can
be represented in terms of products  of  their bilinear combination, i.e.   the
Killing metric $g_{\a\b}$,
\eqn\rrr{ C^{(nm)}_{\a\b} \sim g_{\a\a'} \eta^{\a'\g'} ...g_{\s\s' } ...
\eta^{\d'\b'}g_{\b\b'} \ ,   }
where the number of  $g_{\a\b}$ factors  is   $n$
and the number of $\eta^{\a\b}$-factors (equal to the power of the Planck
constant)
is  $n-1$.
The known two-loop result  \refs{\casw, \ichi}
then implies that
\eqn\rre{  C^{(22)}_{\a\b} =0\ , \ \ \
  C^{(21)}_{\a \b} = \b_2  g_{\a\a'} \eta^{\a'\b'}g_{\b'\b } \ , \ \ \
\b_2= - {17\ov 6 (4\pi)^4 } \ , }
or, equivalently (cf. \reno)
\eqn\renom{ \hat \O_{ab}^{(2-loop)}
= \O_{ab}  + {1 \ov  \ep }\b_1 g_{ab}  +
{1\ov \ep} \b_2 g_{aa'} \O^{a'b'}g_{b' b } \ . }
In principle, it could happen   that in  the  general case
of  non-semisimple Lie algebra $\cal L$
the  matrix $g_{aa'} \O^{a'b'}g_{b' b }$ would be non-vanishing and  not
proportional to $\g_{ab}$ (cf. \kill,\defqe),  i.e.  that starting with the
2-loop level it  would  be necessary
to renormalize also
 $\O^{(0)}_{ab}$. This  would be  strange  since  then it  would
not be  manifest  that  the resulting theory is  renormalizable,
while it must  be such (on symmetry and dimensional consideration grounds)
  provided $\O_{ab}$ was chosen in the most general possible form.

 It is very likely,  however,
that the  very special   double extension structure of  indecomposable algebras
 $\cal L$  \figu\
actually implies
that the conclusions for any of  such $\cal L$
will always be the same
as  for  the  `depth 1' special case
$\cal L$ is a  double extension of an {\it abelian}
$\cal G$   by  a simple  or 1-dimensional  $\cal H$.
In this  case  (which includes all the explicit examples in
\refs{\napwi,\oliv,\sfetso,\sf,\keha})
we  have according to \defe,\defr\
 $$g_{aa'} \O^{a'b'}g_{b' b }
= \pmatrix{0 & 0 & 0 \cr
0 & g_{rs} & 0 \cr
0 & 0 & 0\cr}\   \pmatrix{\O^{(0)ij} & 0 & 0 \cr
0 & 0 & \delta^r_s \cr
0 & \delta^s_r  & - \O_{rs} \cr}\      \pmatrix{0 & 0 & 0 \cr
0 & g_{rs} & 0 \cr
0 & 0 & 0\cr}\ =0\ , $$
\eqn\fefr{ g_{aa'} \O^{a'b'}g_{b' c }...g_{dd'} \O^{d'e'}g_{e' b }=0 \ ,   }
i.e., as in WZW and CS theories,  there is {\it no renormalization}
of  $\O_{ab}$ beyond one loop!\foot{In WZW and CS theories this conclusion
holds (in a properly chosen scheme)
 not only in the non-semisimple but also in the semisimple algebra case.}
Moreover, as we shall show in the next section,  in the case of the `depth 1'
algebras the
full effective action is exactly given by the 1-loop term.

The non-semisimple
YM theories  thus occupy   an `intermediate'
place  between  the  abelian  theories
 (where there is no renormalization at all) and  the semisimple non-abelian
YM  theories  (where there is a renormalization at each  loop order).
It is interesting to note  that in the non-semisimple  theory discussed above
there are  no higher loop  divergences in spite of the fact that
the subalgebra  $\cal H$  of the  full gauge algebra  $\cal L$
may be non-abelian.
This  is due to a   peculiar way in which $\cal H$ is embedded into
$\cal L$. In fact, the non-abelian nature of $\cal H$ is
only  indirectly reflected  even   in the 1-loop renormalization \reno:
the  $\cal H$ part of the 1-loop  divergence \onel\
is proportional to $g'_{rs}$ (defined in \kill)  which
is different  from the Killing metric  $g_{rs}$ of $\cal H$
 appearing  in the 1-loop counterterm of $\cal H$ YM theory  and
is nonvanishing
even if $\cal H$ is abelian (see Section 5).

\subsec{On-shell finiteness and  exact effective action}
The  special structure of non-semisimple theories leads to the conclusion
 that  in spite of the presence of the (1-loop)  divergence in the off-shell
effective action,  they are actually   UV finite on-shell,
i.e. have finite scattering amplitudes.
  To  demonstrate  this
let us consider   the structure of the
divergence \onel\ in more detail.
First,  let us write down explicitly  the YM action \yyym\ in the case of
 generic indecomposable non-semisimple algebra \me.
Introducing the ${\cal G, H , H}^*$ components of the gauge potential  and
computing the field strengths
\eqn\pot{ A^a_\m e_a = A^i_\m e_i + A^r_\m e_r + A_{s\m} e^s\ ,
\ \ \
 F^i_{\m\n} =
2\del_{[\m } A^i_{\n ]}  + f^i_{\ jk} A^j_\m A^k_\n  - 2f^i_{\ jr}  A^r_{[\m }
A^j_{\n ]} \ ,  }
$$ F^r_{\m\n} =
2\del_{[\m } A^r_{\n ]}  + f^r_{\ st} A^s_\m A^t_\n\ ,
 \ \ \
F_{s\m\n} =
2\del_{[\m } A_{s\n ]} +  2f^t_{\ sr}  A^r_{[\m } A_{t\n ]} +  f_{i js}
A^i_{\m } A^j_{\n }  \ , $$
we  find for the YM action (see \defqe)
\eqn\ymeew { S= {1\ov 4} \int d^4x \big( \O_{ij} F^i_{\m\n}  F^j_{\m\n}
 + 2g_{ir} F^i_{\m\n}  F^r_{\m\n}  + \O_{rs} F^r_{\m\n}  F^s_{\m\n}
+  2  F^r_{\m\n}F_{r\m\n} \big) \    }
$$= {1\ov 4} \int d^4x \big( \O_{ij} F^i_{\m\n}  F^j_{\m\n}
 + 2g_{ir} F^i_{\m\n}  F^r_{\m\n}
+  2   F^r_{\m\n}\hat F_{r\m\n}\big) \   , $$
where we have  simplified the action  by introducing the  new  variable
\eqn\sfff{ \hat  A_{s\m} =  A_{s\m} + \ha \O_{sr}  A^r_{\m}\ , \  \
\ \  \hat  F_{s\m\n} =  F_{s \m\n} + \ha \O_{sr} F^r_{\m\n} \ . }
This is possible, in particular, due to the fact that
  the field component $A_{s\m}$ appears in the action only linearly.
Thus it plays the role of a Lagrange multiplier which,  when  integrated out in
the path integral,  gives the factor $\delta ({ D}_\m F^r_{\m\n})$,
i.e. the constraint that $A^r_\m$ should satisfy the classical equations of
motion  of  YM theory  for the algebra  $\cal H$.
As a result, the only dynamical fields that propagate in quantum loops are
$\cal G$-components $A^i_\m$.  Since the structure of $\cal G$
 is strongly constrained by the requirement that $\cal L$ is indecomposable,
the resulting structure of the quantum effective action is  also
very special.

To see this
let us further  specify the discussion to the explicitly tractable
case  when the algebra  $\cal L$ has `depth 1', i.e.  when $\cal G$ is abelian.
Then $f^k_{\ ij}=0$,  the Killing form has only one ($g'_{rs}$) non-vanishing
entry  (see \defe) and the action \ymeew\ takes the form
\eqn\eew { S= {1\ov 4} \int d^4x \big[\  \O^{(0)}_{ij}
\big(2\del_{[\m } A^i_{\n ]}    - 2f^i_{\ lr}  A^r_{[\m } A^l_{\n ]}\big)
\big(2\del_{\m } A^j_{\n }    - 2f^j_{\ kr}  A^r_{\m } A^k_{\n }\big)
} $$ +\  2 \big(2\del_{[\m } A^r_{\n ]}  + f^r_{\ st} A^s_\m A^t_\n\ \big)
\big(2\del_{\m } \hat A_{r\n } +  2f^t_{\ rs}  A^s_{\m } \hat A_{t\n } +  f_{i
jr}  A^i_{\m } A^j_{\n }\big)\big] \ . $$
The above field  redefinition \sfff\  eliminated the dependence of the action
on  $\g$, i.e.  $\g$ (in fact, the whole $\O_{rs}$ in \defe)
is not an `essential  coupling'  \wein\ of the theory.
Since we have seen that the  divergence \onel\ can
be  absorbed into a shift of   $\g$ \reno,  that means that the theory
does not  need a  non-trivial renormalization at all,
having  finite S-matrix elements, i.e.  being    {\it UV finite on shell}.

As it is clear  from \eew,  the action
is bilinear in $A^i_\m$.
Combining this  with the fact that the
path integral over $\hat A_{r\m}$ `kills' the quantum fluctuations of $A^r_\m$
it  is easy to show  that the full
effective action is exactly given by the { \it 1-loop term}
which is  a combination of logarithms of  determinants
depending only on the  background field $A^r_\m$,
\eqn\edd{ \Gamma= S[A^i_\m,A^r_\m,\hat A_{r\m}]  + \Gamma^{(1-loop)}[A^r_\m] \
. \  }
The  divergent part  \onel\
$$
\G^{(1-loop)}_\infty= {1 \ov  \ep }\b_1  g'_{rs}  \int d^4x \   F^r_{\m\n}
F^s_{\m\n} \ , $$
 as expected,   can be eliminated by  the additional (infinite) field
redefinition
\eqn\fie{ \hat  A^{(1-loop)}_{s\m} =  \hat A_{s\m} + {2\ov \ep} \b_1  g'_{sr}
A^r_{\m} \ . }
The
 theory has  effectively
reduced  to that of     abelian vector fields $A^i_\m$
coupled  to the  external (classical)   gauge field $A^r_\m$ corresponding to
$\cal H$.
 This  interpretation  makes it clear that
 though this model does not have
non-trivial UV divergences (and therefore  is
 scale invariant on shell)
 it is not
conformally invariant in the usual sense.
Assuming that the  theory  was  first defined on a curved 4-space, \onel\
 implies
that the conformal factor of the 4-metric does not decouple, i.e.
the expectation value of the trace of the stress-energy  tensor
is
\eqn\anom{  <T^\m_\m> =\b_1 g'_{rs} F^r_{\m\n} F^s_{\m\n}  \ .  }
This  is  an `external'   conformal anomaly,
which is not reflected in a scale dependence of on-shell
correlation functions but
present  in the effective action.
Analogous  anomaly appears, for example,
 in
the theory of quantized interacting  Higgs field coupled to an
external  $U(1)$ gauge field,  where   again it   is  not reflected in the
Higgs field  S-matrix since the gauge field is  classical.


Since the action \eew\
 of the `depth 1'  YM theories
 is linear in the field $\hat A_{\r\m}$
  which can  thus be explicitly integrated out
from the action,  one may hope to be able  to
 define a unitary subsector of  theory
in terms of the resulting `reduced'  action
involving only positive norm fields $A^i_\m$.
 We shall  return
to the discussion of the unitarity issue
at the end of Section 5.

\newsec{Simplest examples:  WZW, CS and YM theories for  $E^c_2$  algebra}

In  this section
we shall consider in detail the  first  non-trivial case of
a non-semisimple  algebra with a nondegenerate invariant form --
the central extension of the euclidean algebra in 2 dimensions $E^c_2$ \jac.
We  will    illustrate  explicitly the general
observations  made  above
and    make clear the analogy between non-semisimple
WZW, CS and YM models.

The algebra  $E^c_2$ is defined  as
\eqn\comm{ [e_3, e_i] = \ep_{ij} e_j \ , \ \  \ [e_i, e_j] = \ep_{ij} e_4 \ , \
\ \
[e_4, e_i] = [e_4, e_3]=0 \ , \ \ \  i,j=1,2 \ , }
where $e_i$ and $e_3$  generate  the  two translations and  the rotation
and $e_4$ belongs to the center.
This is   the simplest example of the general case
\me\ with  $e_i$ as  the generators of  $\cal G$, $e_3=e_r$ as  the generator
of $\eh$ and $e_4=e^r$ as the generator  of $\eh^*$.
The algebra \comm\
has the straightforward $N= 2M + 2$ dimensional generalization \sf\
obtained by replacing the 2-dimensional   $\cal G$
by $M$-dimensional abelian one, i.e. by  adding an extra index  ($I=1,...,M$)
to  $e_i$
and replacing $[e_i, e_j] = \ep_{ij} e_4$
by $[e_{iI}, e_{jJ}] = \delta_{IJ} \ep_{ij} e_4$.
This is  a  `depth 1'  non-semisimple algebra
with a non-degenerate  invariant form  with  only  one minus sign
in the signature (see Section 2).
All of the discussion of  $E^c_2$ theories below can be straightforwardly
generalized to the case of arbitrary  $M >1 $.

\def \bg {\bar \g}

The algebra  \comm\  may be compared to the algebra of  $SU(2) \times U(1)$
\eqn\cmmm{ [{\tilde e}_3, {\tilde e}_i] = \ep_{ij} {\tilde e}_j \ , \ \  \
[{\tilde e}_i, {\tilde e}_j] = \ep_{ij} {\tilde e}_3 \ , \ \ \
[{\tilde e}_4, {\tilde e}_i] = [{\tilde e}_4, {\tilde e}_3]=0 \ , \ \ \
i,j=1,2 \ , }
where $({\tilde e}_i,{\tilde e}_3)$ are the generators of $SU(2)$  and ${\tilde
e}_4$ is the generator of  $U(1)$. The two algebras are related, e.g.,  by the
following limit:
\eqn\limi{{\tilde e}_i= \l\inv  e_i\ , \ \ \ {\tilde e}_3 = e_3  + \l^{-2} e_4\
,  \ \ \  {\tilde e}_4= e_4\ , \  \ \ \l \to  0 \  .  }
The algebra \comm\
has  the  degenerate Killing form (cf. \kill)
but
 admits a non-degenerate  invariant bilinear  form of signature $(+++-)$  \jac\
(cf. \invm,\defqe; $a,b=1,... ,4$)
 \eqn\form{ \Omega_{ab}=k\pmatrix{1 & 0 & 0 & 0\cr
0 & 1 & 0 & 0\cr
0 & 0 & 2\bg & 1\cr
0 & 0 & 1 & 0\cr}\
  ,  \ \ \ \ g_{ab}=\pmatrix{0 & 0 & 0 & 0\cr
0 & 0 & 0 & 0\cr
0 & 0 & 2 & 0\cr
0 & 0 & 0 & 0\cr}\
  ,  \  \  \ \ \bg \equiv  k\inv {\g  }\ .   }
where $\g$ is an arbitrary constant. As we have already  mentioned  above,
 $\g$  is  not be   an essential parameter of  the corresponding  field
theories constructed using \form\
since
the  dependence on it
 can be eliminated from the actions  by a field redefinition.
However,  since the  $\g$-term
 is  effectively generated  at the 1-loop level
in the off-shell
effective action  it is natural to  keep  $\g$
 also in  $\O_{ab}$ which defines the classical action,  interpreting  the
quantum correction  as its  (off-shell) renormalization.

As for all `depth 1' algebras, the value of the  overall scale   $k$  of
$\O_{ab}$
in \wzz,\cher,\yyym\  will also  not be  essential
 -- it  will be possible to   set  it equal to 1 (or make it, e.g.,
arbitrarily small)
by a  rescaling of
the fields. This is certainly  different from what happens   in  semisimple
theories
which nontrivially  depend on level $k$ or YM coupling.

\subsec{$E^c_2$ WZW model}
 It is useful first to  consider  the $E^c_2$ WZW model
\napwi\ since there will be many parallels
with subsequent discussion of CS and  YM theories.
Using the  parametrization
$g=\exp (x_i e_i)$ $ \exp (ue_3 + v e_4)$
the $E^c_2$ WZW action \wzz,\deff\ can be put into the form \napwi\foot{We
ignore the  total derivative term and use the following notation:
$d^2 \s = d\s_1d\s_2, \ \del = \ha (\del_1 -i \del_2) , \ \bd = \ha (\del_1 +i
\del_2)$. }
\eqn\wzw{ I_{E^c_2} = {k\ov 4\pi} \int d^2 \s (\del x_i \bd x_i
+ \ep_{ij} x_j \del x_i \bd u  + 2\bg \del u \bd u + 2\del u \bd v ) \
 }
$$ = {k\ov 4\pi} \int d^2 \s  (\del x_i \bd x_i
+ \ep_{ij} x_j \del x_i \bd u  + 2\del u \bd \hat v ) \ , \ \ \ \hat v = v +
\bg u \ .
 $$
Note that the value of $k$ can be changed
by rescaling the coordinates since  the action is invariant under
\eqn\iii{ x_i \to  s x_i\ , \  \ \ u \to u\  ,  \ \ \ \hat v\to s^2 \hat v \ ,\
\ \   k \to s^{-2} k \ .  }
That means that coordinate-invariant observables
(e.g., the  central charge) will not
 depend on $k$.
Transforming the coordinates $x_1\to x_1 + x_2 \cos u , \
\ x_2 \to x_2 \cos u , \ \ {\hat v } \to {\hat v } + \ha x_1 x_2 \cos u,  $
the action can be  represented  also in the  `plane-wave' form \napwi\
\eqn\acd{I_{E^c_2}  ={k\ov \pi} \int d^2 \s \
( \del x_1 \bd x_1
+\del x_2 \bd x_2 + 2 \cos u\ \del x_1 \bd x_2+  2\del u \bd\hat  v ) \ . }
\acd\  is closely related  \sfetso\
 to the action of the  $SU(2) \times U(1)_{-} $ WZW model (the index  `$-$' is
used to indicate that the $U(1)$ term appears in the action with the opposite
sign).
The latter  is given by  (in the $SU(2)$ parametrization
$g= \exp ({\ha i {\s_1} \th_L})\exp ({\ha i {\s_3} \phi}) \exp ({\ha i {\s_1}
\th_R})$)
\eqn\acts{I_{ SU(2) \times U(1)_{-} }={{\tilde k}\ov 4 \pi} \int d^2 \s \ \big(
 \del \th_L \bd \th_L
+ \del \th_R \bd \th_R  + 2 \cos \phi\ \del\th_L \bd \th_R+ \del \phi \bd \phi
 -\del t \bd t
\big) \ ,}
where $t$ is the coordinate corresponding to the $U(1)$ factor.\foot{Note that
this action does not have an invariance similar to \iii, i.e.
it  nontrivially depends on ${\tilde k}$.}
Ignoring the periodicity  of coordinates in \acts\   we  may set  (cf. \limi)
\eqn\corr{ \th_L= \l x_1\ , \ \    \th_R= \l x_2\ , \ \  \ \  \p= u + \l^2 \hat
v \ , \ \ \  t=u \ , \ \  \
   \     {\tilde k} = \l^{-2} k  \ ,
 \  \  \ \l \to  0 \  ,   }
and then \acts\ goes into \acd.

The special `null' structure of  \wzw\ implies that there  are  no higher
(than one) loop
corrections: the path integral over ${\hat v }$ constrains $u$ to its classical
values
and what remains is a gaussian integral over $x_i$ (equivalently,
the only  1-PI  Feynman diagrams that contribute to  the effective action  are
the one-loop ones  with internal
$x_i$-propagators and  $u,v$-lines as  external ones \napwi). In addition,
because of the  chiral  form  of the interaction term in \wzw\
 (typical to WZW models)
 there  are no
1-loop divergences, i.e. the model is conformal at the quantum level.
This is also  consistent with  its relation  through
\corr\ to the $SU(2)\times U(1)_-$
WZW model (note  that since in the limit ${\tilde k} \to \infty$ the central
charge
has free-theory  value, $c=4$).
Furthermore, the  full quantum effective action
is given by the 1-loop  term ($x_i$-determinant)
depending only on $u$
\eqn\effw{\Gamma_{E^c_2}
= I_{E^c_2} + \ha \log \det \big[ ( \delta_{ij} \bd  + \ep_{ij} \bd u) \del
\big] \ . }
The  `on-shell' ($\del\bd u=0$) part of this determinant can be computed
explicitly.
Introducing $x= x_1 +ix_2, \ x^*= x_1 -ix_2$ one can
 set $x= e^{iu} y, \ x^* = e^{-iu} y^*$.
When $\del\bd u=0$ the resulting determinant becomes
$u$-independent up to the anomaly contribution which
(in the  left-right symmetric  scheme)
    is given by\foot{This expression corresponds to the 2-sphere.
It is easy also to compute  the determinant  on the 2-torus obtaining the
partition function  \kk\ of this model \rts.}
  $$ {1\ov 2\pi}   \int d^2 \s   \del u \bd u \ . $$
 While there is no shift of the parameter  $k$,  as expected,
the constant  $\g$  gets  the    quantum correction
\eqn\quac{  \hat {\bg}=   \bg +  {1\ov k} \  , \ \  \ \  \ \ \ \hat \g=\g + 1 \
. }
This  agrees with the general result \mass,\defq\
(the same one-loop shift was  found in
\refs{\napwi,\kk}).
The  general
`off-shell' expression for the effective action for  an arbitrary $u$
contains also  non-local terms which depend on $\del\bd u$.

The simplicity of this  theory is  due  to the  special structure of the action
which
 is linear in $v$, quadratic in $x^i$ and has chiral  interaction term.
As a consequence,
can explicitly solve the corresponding   field equations
(and thus the  resulting conformal field theory)
in terms of free fields (the dependence on on-shell values of $u$ can be
effectively `eliminated'    by  the `rotation' of $x^i$).
In spite of the existence of one time-like direction in the field space,
the  CFT  corresponding to \wzw\  is unitary
since (in contrast to, e.g.,  $SL(2,R) $ WZW model) here  one can use  the
light-cone gauge \refs{\kk,\rts,\palla}.

\subsec{$E^c_2$ Chern-Simons theory }
 The Chern-Simons action \cher\  for $E^c_2$ has the following explicit
structure
\eqn\chr{ S_{E^c_2}
= - { ik\ov 8\pi}  \int  d^3 x \  \ep^{\m\n\l}
\big( A^i_\m \del_\n  A^i_\l    +  \ep_{ij}  A^i_\m A^j_\n A^3_\l
+ 2\bg A^3_\m  \del_\n   A^3_\l  +   2A^3_\m\del_\n  A^4_\l  \big) \  }
$$ = \ - { ik\ov 8\pi}  \int  d^3 x \  \ep^{\m\n\l}
\big( A^i_\m \del_\n  A^i_\l    +  \ep_{ij}  A^i_\m A^j_\n A^3_\l
  +   2A^3_\m\del_\n  {\hat A}^4_\l  \big) \ , \ \ \ {\hat A}^4_\l
= { A}^4_\l +  \bg A^3_\l \ ,  $$
which is closely  related to that of \wzw.
As in WZW case, $k$ can be  changed by rescaling $A^i_\m, {\hat A}^4_\m $
and \chr\ can be  also obtained   as a limit of the semisimple  $SU(2)\times
U(1)_-$
Chern-Simons theory (the derivation is similar to the one discussed below
in the YM case).
Since the integral over $\hat A^4_\l$
implies that $A^3_\m$ is  subject to the classical equation,
 the quantum part of the effective action depends only on  $A^3_\m$ and is
exactly given by the
one-loop term.  The local `on-shell'
part of the
determinant resulting from integration  over  $A^i_\m$
can be explicitly computed  (see, e.g., \refs{\ferm, \shif})
so that the final result is (cf. \cherr,\chee, \effw)
\eqn\quach{ \Gamma_{E^c_2} = S_{E^c_2}
+ \ha \log \det \big[  \ep^{\m\n\l} (\delta_{ij}  \del_\m     +  \ep_{ij}
A^3_\m) \big]   }
$$ = S_{E^c_2}     -    { i \ov 4 \pi}  \int  d^3 x \  \ep^{\m\n\l}
A^3_\m  \del_\n A^3_\l  + ...  \ ,  $$
where dots stand for non-local $P$-even  `off-shell' terms which
vanish when  $\del_{[\m} A^3_{\n]} \to  0$.
Again, in contrast to the semisimple case  \refs{\wiit,\others,\shif}),  there
is no quantum renormalization of $k$ but  only
a shift of the coefficient $\g$  which is the same as in
 $E^c_2$ WZW theory \quac.

\subsec{$E^c_2$ Yang-Mills theory }
Starting with \comm,\form\  one finds from \yyym\
  (cf. \chr)
\eqn\yme { S_{E^c_2}= {1\ov 4g^2} \int d^4x \big( F^i_{\m\n}  F^i_{\m\n}
 + 2\bg F^3_{\m\n}  F^3_{\m\n} +  2 F^3_{\m\n}  F^4_{\m\n}\big) \  }
$$ = \  {1\ov 4g^2} \int d^4x \big( F^i_{\m\n}  F^i_{\m\n}
 +  2 F^3_{\m\n}  \hat F^4_{\m\n}\big) \ , $$ \eqn\fff{  F^i_{\m\n} =
2\del_{[\m } A^i_{\n ]}  - 2\ep_{ij}  A^3_{[\m } A^j_{\n ]}
\ , \ \ \ \ \ \  i=1,2 \ , } $$  \ \  \ F^3_{\m\n} =
2\del_{[\m } A^3_{\n ]}\ ,
\ \  \  \ \
F^4_{\m\n} =
2\del_{[\m } A^4_{\n ]} + \ep_{ij}  A^i_{\m } A^j_{\n }  \ ,     $$
$$
  {\hat A}^4_\m
= { A}^4_\m + \bg A^3_\m \ , \ \ \ \  g^2=1/k \ . $$
This is a  special case of the YM action for `depth 1' non-semisimple algebras
\eew\
(here $A^r_\m = A^3_\m, \ A_{r\m}= A^4_\m, \ \ f^r_{\ st}=0,\ f_{ijr} =\ep_{ij}
$).
It is invariant under \gau,
\eqn\inva{ \delta A^i_\m = \del_\m \eta^i + \ep_{ij } A^j_\m \eta^3
- \ep_{ ij } A^3_\m \eta^j\ ,  \ \ \
\delta A^3_\m = \del_\m \eta^3 \ , \ \  \
\delta A^4_\m = \del_\m \eta^4 + \ep_{ij}  A^i_\m \eta^j\ .}
The correspondence  between the YM and   WZW    actions \yme\ and \wzw\
constructed using the same algebra and  invariant form is established by
 \eqn\ouo{ x_i \sim  A_\m^i\ , \ \ \  u\sim  A^3_\m\ , \ \ \ v \sim  A^4_\m\ ,
\ \ \ k\sim  g^{-2} \ .}
 Like \wzw\ the action \yme\ is  quadratic in $A^i_\m$
and the interaction term involves only $A^3_\m$ and $A^i_\m$.
This action  is also invariant  under (cf. \iii)
\eqn\jjj{ A^i_\m \to  s  A^i_\m \ , \  \ \ A^3_\m  \to  A^3_\m \  ,  \ \ \ \hat
A^4 \to s^2  \hat A^4_\m  \ ,\ \ \   g \to s g  \ ,   }
 implying that the theory can only have  trivial dependence
on $g$.  In particular, as in the case of WZW and CS theories,
the 1-loop approximation should  be exact. The 1-loop term
can only  depend  on $A^3_\m$ (cf. \chr)
which is not transformed under \jjj.

Like the WZW action  \wzw,\acd\ the YM action \yme\ can be also obtained as a
limit
of   the  semisimple $SU(2)\times U(1)_-$ YM action
 \eqn\ymee { S_{SU(2)\times U(1)_-} = {1\ov 4{{\tilde g}}^2} \int d^4x \big({
\F}^i_{\m\n}  {\F}^i_{\m\n}
 + \F^3_{\m\n}  \F^3_{\m\n} -  \F^4_{\m\n}  \F^4_{\m\n} \big)\ ,  }
\eqn\ff{  \F^i_{\m\n} =
2\del_{[\m } \A^i_{\n ]}  - 2\ep_{ij}  \A^3_{[\m } \A^j_{\n ]}
\ , \ \ }
$$ \ \F^3_{\m\n} =
2\del_{[\m } \A^3_{\n ]}+\ep_{ij}  \A^i_{\m } \A^j_{\n } \ ,
\ \  \  \ \ \
\F^4_{\m\n} =
2\del_{[\m } \A^4_{\n ]}  \ ,   $$
where  $\A^i_\m$, $ \A^3_\m$ are the $SU(2)$ fields and $\A^4_\m$
is  the `ghost' $U(1)$ field
 (cf. \cmmm,\comm,\fff).
Introducing  the new  variables and  rescaling the coupling constant
(cf. \corr)
\eqn\corre{ \A^i_\m = \l A^i_\m \ ,  \  \ \ \   \A^3=A^3_\m  + \l^2 \hat A^4_\m
\ ,   \ \
     \ \ \A^4_\m = A^3_\m  \ ,  \ \  \ \ {\tilde g} = \l g  \ ,
 }
we can rewrite the   ${SU(2)\times U(1)_-}$ YM  action \ymee\ as
 \eqn\ymeee {  S_{SU(2)\times U(1)_-}= {1\ov 4{g}^2} \int d^4x
\big[  \big (F^i_{\m\n}
- 2\l^2 \ep_{ij}  A^4_{[\m } A^j_{\n ]}
\big)^2
+ ( 2 F^3_{\m\n}  + \l^2 \hat F^4_{\m\n})\hat F^4_{\m\n} \big] }
$$ = \  S_{E^c_2} +  O(\l^2)  \ ,   $$
where we have  used $E^c_2$ field strengths $F^a_{\m\n}$
 given  in \fff.  Thus $S_{E^c_2}$ is the $\l \to 0$  limit of $S_{SU(2)\times
U(1)_-}$.

The quantum theory defined by the $E^c_2$ action \yme\  is much simpler than
the $SU(2)\times U(1)_-$  one
of which it is a limit.\foot{Note, in particular,  that  the norm of the $\hat
A^4_\m$ field becomes null
in the  limit.
Eq. \ymeee\  implies  certain
relations  between   the two quantum theories
(the redefinition  \corre\ is nondegenerate for any finite $\l$).
For example, the vacuum partition function
of $E^c_2$ YM theory $Z(g)$  equal to the $\l\to 0$
 limit of $Z_{SU(2)\times U(1)_-}( {\tilde g})$
must  be trivial
 since in this limit ${\tilde g} \to 0$.}
Consider the path integral corresponding to \yme
\eqn\qua{ Z= \int [dA^i_\m][dA^3_\m][d\hat A^4_\m]
\exp \big( - {1\ov  g^2} \int d^4x
\big[  (\del_{[\m } A^i_{\n ]}  - \ep_{ij}  A^3_{[\m } A^j_{\n ]}
)^2}
$$
+  \ \del_{[\m } A^3_{\n ]}
 ( 2 \del_{\m } \hat A^4_{\n } +   \ep_{ij}  A^i_{\m } A^j_{\n }
 )
\big] \big) \ . $$
 The integral over $(\hat A^4_\m)^{\bot}$
gives\foot{One can use, e.g.,  the gauge
$\del_\m A^a_\m=0$ in which the non-trivial part  $\Delta_{gh}= \det [  \del_\m
(\delta_{ij} \del_\m      - \ep_{ij} A^3_\m)] $ of the
ghost determinant which follows from \inva\
depends only on $A^3_\m$.}
 $\delta (\del_\m F^3_{\m\n}) $, i.e.
fixes  $A^3_\m$
to  be equal to its classical value.  In the vacuum case ($A^3_\m =0$)
 what remains is  the free integral over $A^i_\m$.
Since the 1-PI Feynman diagrams in this theory can only
have   $A^i_\m$ (and their ghosts) propagating in  the
 internal lines (with the  external
lines being represented only by the  $A^3_\m,\hat A^4_\m$-propagators),
the  possible loop diagrams are 1-loop ones only, i.e.
 the  quantum effective action in this theory
 is exactly given by the one-loop term.

The latter can be found  by treating $A_\m^3$ as a
background field and integrating over $A^i_\m$,
\eqn\gamm{ \exp (-\Gamma^{(quant)}_{E^c_2} [A^3_\m]) }
$$ =\int [dA^i_\m]\
\exp\big( - {1\ov {g}^2} \int d^4x
\big[  (\del_{[\m } A^i_{\n ]}  - \ep_{ij}  A^3_{[\m } A^j_{\n ]}
)^2
+  \del_{[\m } A^3_{\n ]} \ep_{ij}  A^i_{\m } A^j_{\n }
\big] \big) \ . $$
It is interesting to note that
$\Gamma^{(quant)}_{E^c_2} [A^3_\m]$ can be expressed in terms of
 the one-loop effective action in $SU(2)$ YM theory.
Suppose one starts with the  $SU(2)$ part of
 \ymee\ and introduces a background only  for  the third component of the
gauge field.
Expanding the $SU(2)$ action to the second order in the quantum fields we get
(as in \onee\ here  $A^3_\m$  is   the  background field
and $B^i_\m, B^3_\m$ are    the quantum fields)
\eqn\yyh { S_{SU(2)} =   {1\ov {g}^2} \int d^4x
\big[\    (\del_{[\m } B^i_{\n ]}  - \ep_{ij}  A^3_{[\m } B^j_{\n ]}
)^2}
$$
+ \   \del_{[\m } A^3_{\n ]} \ep_{ij}  B^i_{\m } B^j_{\n }
+ (\del_{[\m } B^3_{\n ]})^2    + O((B^a_\m)^3)
\big]  \ . $$
Observing that  $B^3_\m$ is decoupled from $A^3_\m$
and comparing with \gamm\
we conclude that  the exact quantum part
of the effective action in $E^c_2$ YM
theory is equal to the one-loop effective
action in $SU(2)$ YM theory with only $A^3_\m$ as its non-vanishing argument,
\eqn\quae{\Gamma_{E^c_2}[A^i_\m, A^3_\m, A^4_\m] = {1\ov 4g^2} \int d^4x \big(
F^i_{\m\n}  F^i_{\m\n}
 +2 F^3_{\m\n}  \hat F^4_{\m\n}\big)  +  \Gamma_{SU(2)}^{(1-loop)}[ A^3_\m] \ .
  }
If $\Gamma_{SU(2)}^{(1-loop)}[A^3_\m]$ is  computed in
the background field gauge \refs{\dewit,\hooft}:
\eqn\gauf{
D_\m B^i_\m \equiv  \del_\m B^i_\m  - \ep_{ij } A^3_\m B^j_\m
 = \xi^i(x)\ , \ \ \ \del_\m B^3_\m = \xi^3(x) \ ,  }
then the effective action $\Gamma_{E^c_2}$ is invariant under the classical
gauge symmetry \inva, i.e.,  $\delta A^3_\m  = \del_\m \eta^3$.
$\Gamma_{SU(2)}^{(1-loop)}$
contains the well-known divergent term \onel, i.e.,
\eqn\div {
(\Gamma_{SU(2)}^{(1-loop)}[A^3_\m])_{\infty}
= {2\ov \ep} \b_1  \int d^4 x  F^3_{\m\n} F^3_{\m\n} \  .  }
As was already discussed above, this divergence
can be  absorbed into the renormalization of $\gamma$
\reno\ or
 can be eliminated by the field
redefinition \fie\ (cf. \fff)
\eqn\renk{ \hat A_\m^{4(1-loop)}=   \hat A^4_\m  + {4 \ov \ep} g^2\b_1  A^3_\m
\ . }
To determine the  `ground state' of the theory one is
 to solve the  effective equations that follow from
$\Gamma_{E^c_2}$.  The equation for $A^3_\m$
is still the classical Maxwell one, the equation for $A^i_\m$ is linear
and  does not receive quantum corrections\foot{The equation for $A^i_\m$
that follows from \yme\   has the same form
as the linearized $SU(2)$ YM equation  in the  $A^3_\m$ background
(i.e. the equation for $B^i_\m$ that follows from \yyh).
Introducing $W_\m = A^1_\m + i A^2_\m, \ W^*_\m = A^1_\m - i A^2_\m,$
one finds: $ - D_\m D_{[\m } W_{\n]} - i F^3_{\m\n} W_\n =0, \
\ D_\m = \del_\m + i A^3_\m . $ }
and   to  find  $A^4_\m$
one needs only to compute the derivative of
$\Gamma_{SU(2)}^{(1-loop)}[A^3_\m]$ at the
classical value of $A^3_\m$.
The  solution of the effective equations  is thus  parametrized   by
 a  solution $(A^3_\m)_*$
of the Maxwell equation.

 As we have already discussed in Section 4.2,
the theory still has  the  `external' conformal anomaly \anom, i.e.
$<T^\m_\m>  = 2\b_1 F^3_{\m\n} F^3_{\m\n}$.
The S-matrix (which is finite and scale invariant)
has the following structure.
 The scattering amplitudes with external $A^3_\m$ points (ends of $<A^3_\m\hat
A^4_\n>$ propagator lines)  vanish,
the amplitudes
with  external $A^i$ lines are   given  by the tree diagrams only,
and the loop correction \quae\ gives
non-trivial scattering  amplitudes with only $\hat A^4_\m$ as external points
(such amplitudes  are  absent  at the tree level).

The fact that the  Euclidean action and Hamiltonian
of this  theory are not positive definite
implies  the
absence of unitarity.\foot{The energy is non-positive since
there are 6  negative norm components  $A^i_0$ $ ,A^+_0,$ $  A^-_\a, \ $ $
(\a=1,2,3),
\ A^{\pm}_\m \equiv  A^3_\m \pm \hat A^4_\m, $ and only 4 gauge  symmetries.}
Still, the  special  structure of the interaction term in \yme\
(which does not involve $\hat A^4_\m$) and of quantum scattering amplitudes
 (with no $A^3_\m$ fields
appearing as  external points)
suggests that there may exist
a re-interpretation of this model which is  consistent with
unitarity.
The relation \ymeee\ to $SU(2)\times U(1)_-$ model \ymee\
seems to make this plausible.
While the $SU(2)\times U(1)_-$  action   \ymee\
is non-positive, the `non-unitarity' of this theory
caused by the negative sign of the $U(1)$ term is a
{ trivial} one since  the $U(1)$ field is completely decoupled:
the  non-trivial part of the full  S-matrix is
just  the unitary $SU(2)$  Yang-Mills S-matrix.
Since the field redefinition \corre\
can not  change this conclusion
for any finite $\l$, one  may hope that the  theory obtained in the limit
 $\l\to 0$ is also unitary.


\newsec{Conclusions}
Our  motivation for considering  non-semisimple Yang-Mills theories
 was their close analogy with non-semisimple
 conformal WZW models in 2 dimensions (and CS models
in 3 dimensions) which have particularly transparent  structure and are
explicitly solvable in terms of free fields.
We have seen that these YM models are indeed
much  simpler  than  the standard  compact  non-abelian theories
and come very close to be 4-dimensional `free-field' analogs of the
2-dimensional models.
They have UV finite yet non-trivial S-matrix
controlled by the effective action containing only  1-loop term.
The main  issue is,  of course,    unitarity. It
may, hopefully,  be resolved
by using the interpretation of these
non-semisimple  models as  special limits of semisimple YM models
which are superficially non-unitary but
have   ghosts  that  are  decoupled from a unitary sector.

\bigskip\bigskip
\newsec{Acknowledgements}
I would like to thank I. Kogan  and K. Sfetsos for  useful discussions.
I  also  acknowledge the  support
of PPARC, EEC grant SC1$^*$-CT92-0789
and NATO grant CRG 940870.

\vfill\eject

\listrefs
\end


To clarify this let us consider a  similar $D=4$ theory for scalar fields
which has the same   apparently non-unitary structure,
\eqn\sca{  L = (\del_\m \vp_1)^2 - (\del_\m \vp_2)^2  + (\del_\m \p)^2
 + L_{int} (\vp_1 - \vp_2, \ \p ) \ , }
where the interaction term depends only on one
linear combination of the positive norm field $\vp_1$ and the ghost field
$\vp_2$.
If instead of $\vp_{1,2}$ we use $\vp_\pm = \vp_1 \pm \vp_2$ to define the
asymptotic states of the theory
then since  $\vp_+$
is decoupled from $\p$
\eqn\scal{  L = \del_\m \vp_+ \del_\m \vp_-  + (\del_\m \p)^2
 + L_{int} (\vp_-, \ \p ) \ . }
It is then natural to demand unitarity only in the $\vp_-,\p$ subsector
of the theory.
Soop, $\vp_-$ and $\vp_+$ do not define propagator on their own ??
To make the analogy with \eew\ even more explicit
let us assume that the action is bilinear in $\p$,
\eqn\scall{  L = \del_\m \vp_+ \del_\m \vp_-  + (\del_\m \p)^2
 + f (\vp_-) \p^2  \ . }